\documentclass[a4paper]{jpconf}
\pdfoutput=1
\usepackage{graphicx}
\usepackage{amsmath}
\usepackage{pstricks}
\usepackage[charter,expert]{mathdesign}

\usepackage{cite}
\usepackage[%
  unicode=true,%
  colorlinks=true,%
  linkcolor=blue,anchorcolor=blue,%
  breaklinks=true,%
  citecolor=blue,filecolor=cyan,%
  menucolor=blue,%
  urlcolor=blue]{hyperref}

\begin{document}
\title{Interaction of a single mode field cavity with the 1D XY model: Energy spectrum}

\author{H Tonchev, A A Donkov, H Chamati}

\address{Institute of Solid State Physics,
Bulgarian Academy of Sciences, \\
72 Tzarigradsko Chauss\'ee, 1784 Sofia, Bulgaria}

\ead{htonchev@issp.bas.bg, aadonkov@issp.bas.bg, hassan.chamati@issp.bas.bg}
%\ead{h\_tonchev@phys.uni-sofia.bg, aadonkov@issp.bas.bg, hassan.chamati@issp.bas.bg}

\begin{abstract}
In this work we use the fundamental in quantum optics Jaynes-Cummings
model to study the response of spin $\tfrac12$ chain to a single mode
of a laser light falling on one of the spins, a focused interaction
model between the light and the spin chain. For the spin-spin
interaction along the chain we use the XY model. We report here the
exact analytical results, obtained with the help of a computer algebra
system, for the energy spectrum in this model for chains of up to 4
spins with nearest neighbors interactions, either for open or cyclic
chain configurations. Varying the sign and magnitude of the spin
exchange coupling relative to the light-spin interaction we have investigated 
both cases of ferromagnetic or antiferromagnetic spin
chains.
\end{abstract}

\section{Introduction}
The Jaynes-Cummings \cite{EJFC1963} model involving the
interaction of an atom with a quantized electromagnetic field describes
electron transitions between the atom levels induced by the modes of
the field. Since its introduction this interaction have found many
applications. It has been suggested that an ion trap of atoms or
ions could be a realization of a quantum computer. In this context the
Jaynes-Cummings model was used in~\cite{WVRM1995} to investigate atom- or
ion-field interactions in an ion trap and in~\cite{JYFN2003} in
quantum dots. Other uses of this model are for one photon
lasers~\cite{GRHW1987} and for single-photon photo
detectors~\cite{ADSM2006}. For some pedagogical texts describing the
model see, for example~\cite{CGPK2005,MSMZ1997}.

In solid state physics there are many models used to
quantum-mechanically describe the interaction between the spins in a spin
chain. Commonly used models are the quantum Ising~\cite{JD2005} and
the Heisenberg~\cite{MASB2001} ones. In the Ising model the spins are
allowed the fewest degrees of freedom -- only one. If the spins have
all the degrees of freedom -- three, then the Heisenberg model is used.
In the intermediate XY model~\cite{JBMC2010} the spins interact
through two degrees of freedom. The XY model is used to study
entanglement~\cite{JBMC2010},
spin glasses~\cite{JMDG1997}, phase transitions~\cite{JMDG1997},
and the W state in quantum
computing~\cite{XW2001}.

In the literature, there exist several models describing the collective interaction
between the electromagnetic field and a spin chain. One such model is
the Dicke model, often used to study the superadiance phase
transitions~\cite{YWFH1973}. Another recent model~\cite{STHA2014}
describes the collective interaction between classical monochromatic
circularly polarized light and the spin chain. Here, we
report a novel use of the Jaynes-Cummings model, namely, the study
of the response of a spin $\tfrac12$ chain to a single quantum mode of a
laser light focused on a particlar spin.

The description of the XY-Jaynes-Cummings model for the interaction
Hamiltonian follows in the next Section~\ref{sec:Hamiltonian}, after
that we report results obtained by solving the model for the particular
case of a finite chain of four spins in
Section~\ref{sec:examples4spins}, followed by concluding remarks.

\section{The Hamiltonian}
\label{sec:Hamiltonian}
\begin{figure}[ht!]\centering
\includegraphics[width = 2.2in]{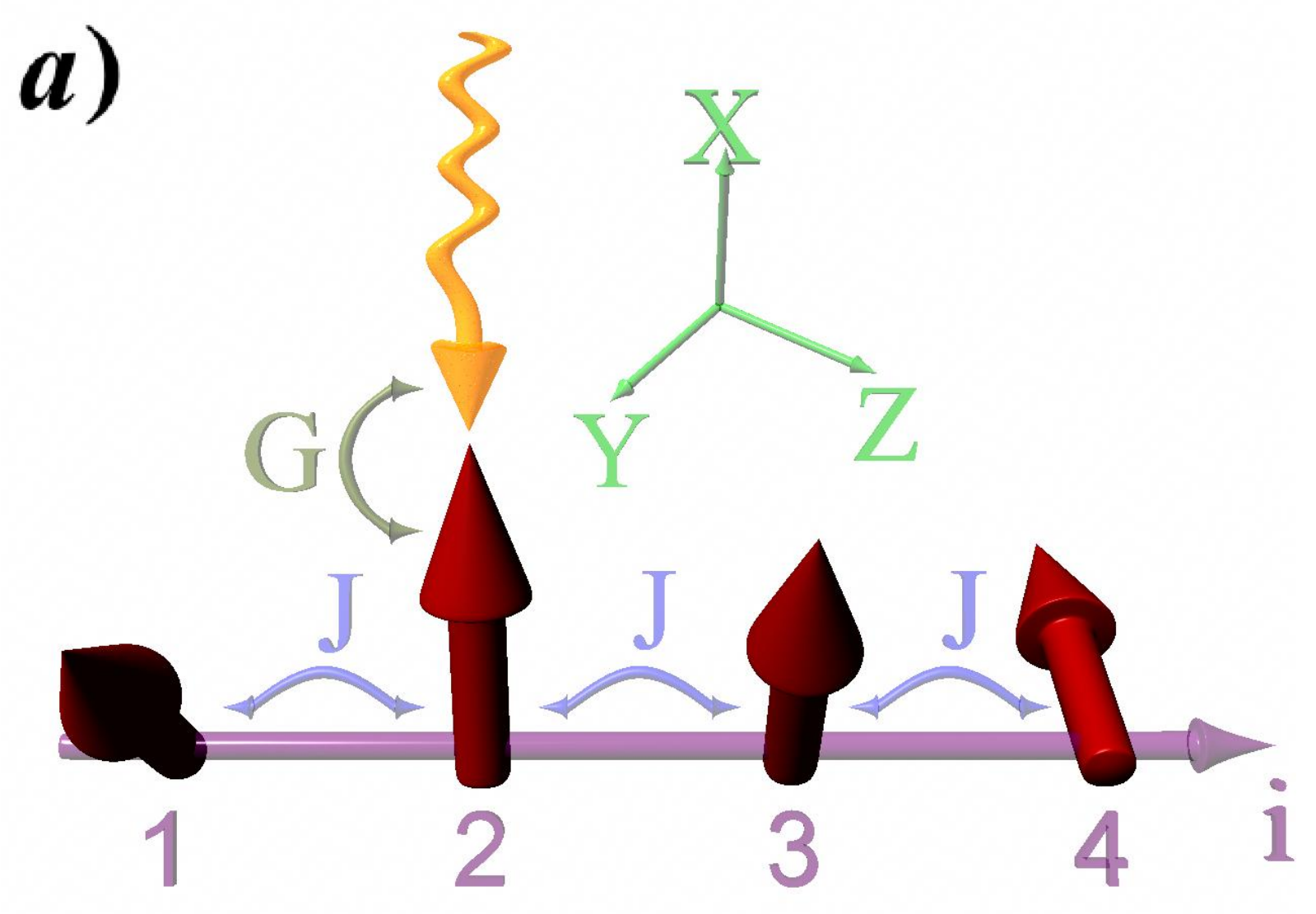}
\hspace{1.0in}
\includegraphics[width = 1.5in]{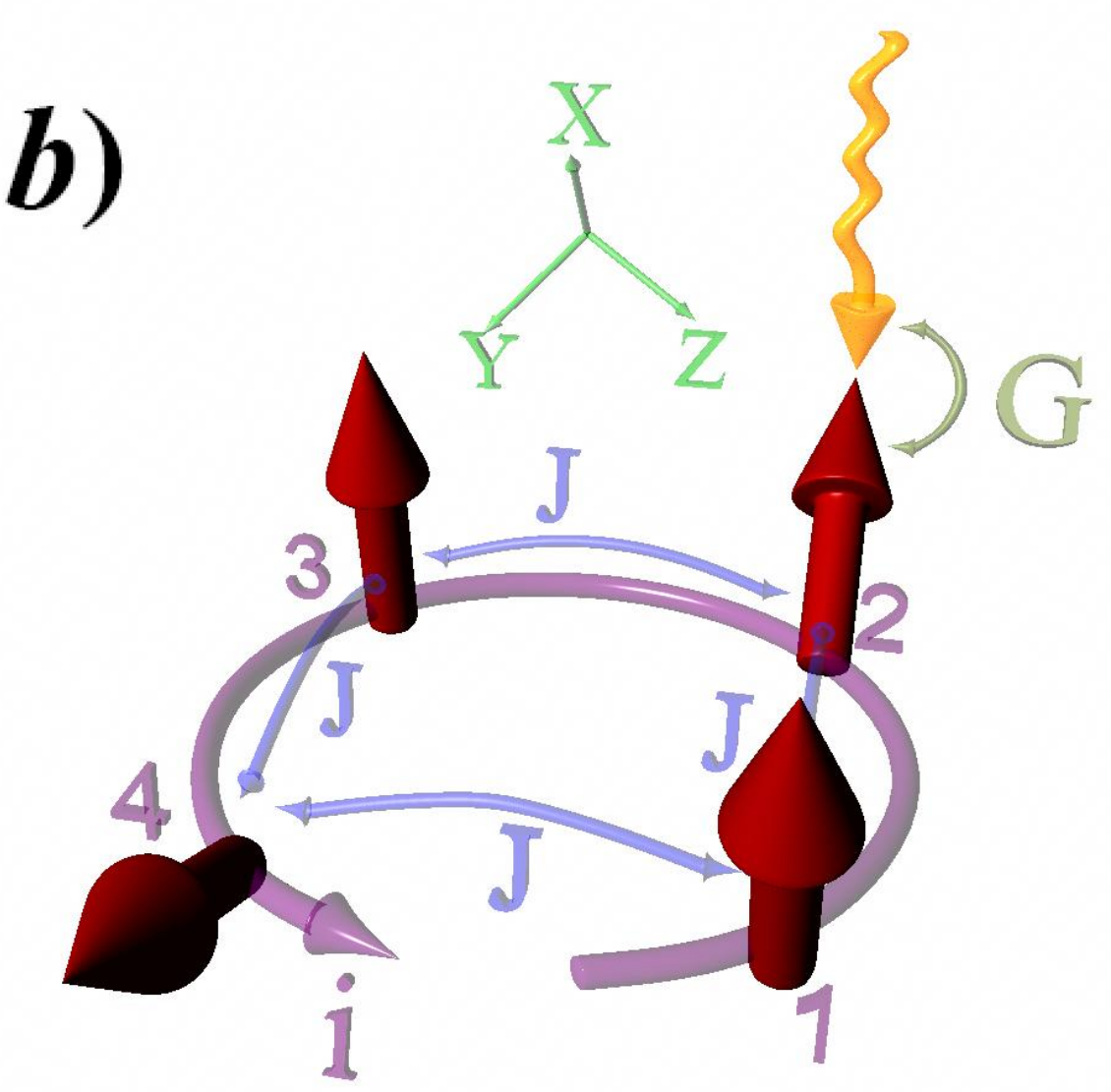}%2.1
\caption{Theorist's rendering of open (a), or cyclic (b)
configurations of four spin chains. The wiggle line represents the
photon.}
\label{Fig:PicturesOfModels}
\end{figure}

The object of our study is a spin $\tfrac12$ chain of $N$ sites and
one mode of a laser field interacting with one of the spins in the
chain. The spins form either an open or a closed, cyclic chain, for
which the last spin interacts with the first, see
Figure~\ref{Fig:PicturesOfModels}. We consider chains with nearest
neighbor interactions $J$, in the XY Model, which couples the $S^x$
and $S^y$ components of the neighboring $i,j$ spins operators
$S_{i,j}=(S_{i,j}^x,S_{i,j}^y,S_{i,j}^z),$ and the field is coupled to
the $k-$th spin.

The interaction Hamiltonian $\widehat{H}$ reads~\cite{remark}:
%$\widehat{H}_{C}$ for cyclic, are:
\begin{equation} 
\widehat{H}= G\left(a^{}S_k^{+} + a^\dag S^-_k \right)-
2J\sum_{i=1}^{N-1}\left(S_i^xS_{i+1}^x+S_i^y S_{i+1}^y\right),
%\;\;\widehat{H}_{C}= \widehat{H}_{O}-2J\left(S_1^x S_{N}^x+ S_1^y S_{N}^y\right).
\label{Eq:HamOCOperator}
\end{equation}
where the first term is nothing but the Jaynes-Cummings model. Here $G$ incorporates
information about the interaction between the mode of the field and
the $k$-th spin, and $S^+$ ($S^-$) lift (lower) the spin projection,
respectively, and are given by $S^\pm = S^x\pm \textrm{i} S^y$, with
``$\textrm{i}$'' the imaginary unit. They would correspond to the
transitions between the lower and the upper atomic levels in the
commonly used formulations of the Jaynes-Cummings model. The operators
$a$ and $a^\dag$ are the annihilation and creation operators of the
mode of the field. The expression with the spin exchange interaction
$J$ between the spin dipoles is the XY model.
%Factor 2 is chosen for convenience.
With the ``$-$'' sign in mind, in the classical limit of
spin vectors, $J>0$ would mimic a mostly ferromagnetic (FM) alignment in
the chains (the $S_z$ components being arbitrary), $J<0$ an
antiferomagnetic (AFM) alignment, and $J=0$ would describe independent spins
along the chain. 
%\textcolor{red}{The diagonal terms in the total Hamiltonian $\sim (a^{\dag}a+\sum\nolimits_i S^{z}_i) $ form an invariant and have been rotated away with a unitary transformation~\cite{RJZSHM2015}.}
%%%%%fig.1%%%%%%%%
%%%%%fig.1%%%%%%%%

For the calculations we choose a basis set of states by first ordering
the field states, ($| 0 \rangle,  | 1 \rangle),$ where $0\textrm{ and
} 1$ stand for the number of photons, then for each spin site we
have the spin states ($|\uparrow \rangle,|\downarrow \rangle$). This
gives a set of $2^{N+1}$ basis states. The spin operators in this
basis are expressed through the Pauli matrices by the standard
relations, $S^{x,y,z}=\tfrac12\hbar \sigma^{x,y,z}$. Since the spin-spin
term $S_i^xS_{i+1}^x+S_i^y S_{i+1}^y$ can be written as
$\tfrac12(S_i^+S_{i+1}^-+S_i^- S_{i+1}^+)$, it useful to define additionally
$\sigma^\pm = \tfrac12\left(\sigma^x \pm \textrm{i} \sigma^y\right)$, so by
definition $S^\pm = \hbar \sigma^{\pm}_{}$. To summarize for spin
$\tfrac12$ we have the expressions:
\begin{subequations}
\label{Eq:PauliMatr}
\begin{equation}
% S^{x,y,z}=\frac{\hbar}{2} \sigma^{x,y,z},\;\;
  \sigma^{x}= 
            \begin{pmatrix}
            0   &   1 \\    
            1   &   0
            \end{pmatrix}, \;\;
 \sigma^{y}= 
            \begin{pmatrix}
            0   & \textrm{-i} \\    
            \textrm{i}   &  0
            \end{pmatrix}, \;\;
 \sigma^{z}= 
            \begin{pmatrix}
            1   &  0 \\    
            0   & -1
            \end{pmatrix}, \;\;
\textrm{ and  } \ \ \
\mathbb{I}= 
            \begin{pmatrix}
            1   &  0 \\    
            0   &  1
            \end{pmatrix},
%\sigma^+= \begin{pmatrix}
%            0   &   1 \\    
%            0   &   0
%            \end{pmatrix}, \;\;  
%\sigma^-= \begin{pmatrix}
%            0   &  0 \\    
%            1   &   0
%            \end{pmatrix},
\end{equation}
\begin{equation}
|\uparrow \rangle = \begin{pmatrix}
            1 \\    
            0
            \end{pmatrix}, \;\;  
|\downarrow \rangle = \begin{pmatrix}
            0 \\    
            1
            \end{pmatrix}, \;\;  
\end{equation}
\end{subequations}
and for the photon we have
\begin{subequations}
\label{Eq:sigmaupdown}
\begin{equation}
 a = \begin{pmatrix}
            0   &   1 \\    
            0   &   0
            \end{pmatrix}, \;\;  
a^{\dag}  =\begin{pmatrix}
            0   &   0 \\    
            1   &   0
            \end{pmatrix}, \;\;
\end{equation} 
\begin{equation}
| 0 \rangle = \begin{pmatrix}
            1 \\    
            0
            \end{pmatrix}, \;\;  
 | 1 \rangle = \begin{pmatrix}
            0 \\    
            1
            \end{pmatrix}.
%a^{\dag} | 0 \rangle = | 1 \rangle,
\end{equation} 
\end{subequations}
% \begin{eqnarray}
% a \textrm{ and } \sigma^+\equiv(\sigma^x+i\sigma^y)/2 = \begin{pmatrix}
%            0   &   1 \\    
%            0   &   0
%            \end{pmatrix}, \;\;  
%a^{+} \textrm{ and } \sigma^-\equiv(\sigma^x-i\sigma^y)/2 =\begin{pmatrix}
%            0   &   0 \\    
%            1   &   0
%            \end{pmatrix},
% \label{Eq:sigmaupdown}
% \end{eqnarray} 
Since the field and the spin operators on different sites act
independently, the matrix form for the subscripted $\sigma_i$ is a
tensor product of $N+1$, $2\times2$ matrices as follows (e.g. $\S$2.5
of~\cite{Rumer:1970}):
\begin{eqnarray} 
\sigma_i^{x}=\underbrace{\mathbb{I}}_\mathrm{Field\
mode}\otimes\underbrace{\mathbb{I}\otimes \mathbb{I}\otimes\cdots\otimes \mathbb{I}}_{i-1 \
\mathrm{terms}}\otimes\underbrace{\sigma^{x}}_{i\mathrm{th\
spin}}\otimes \underbrace{\mathbb{I}\otimes \mathbb{I}\otimes\cdots\otimes \mathbb{I}}_{N - i\ \mathrm{terms}},
\label{Eq:KronekerDefiniion}
\end{eqnarray}
and similarly for $\sigma_i^{y},\sigma_i^\pm.$ This gives for the
field-spin parts of the Hamiltonian:
\begin{eqnarray} 
a^\dag \sigma_j^{y}=\underbrace{a^\dag}_\mathrm{Field\ mode}\otimes\underbrace{\mathbb{I}\otimes \mathbb{I}\otimes...\otimes \mathbb{I}}_{j-1 \
\mathrm{terms}}\otimes\underbrace{\sigma^{y}}_{j\mathrm{th\ spin}}\otimes \underbrace{\mathbb{I}\otimes \mathbb{I}\otimes...\otimes \mathbb{I}}_{N - j\ \mathrm{terms}},
\label{Eq:KronekerField}
\end{eqnarray}
and for the spin-spin terms:
\begin{eqnarray} 
\sigma_i^{x}\sigma_j^{y}=\underbrace{\mathbb{I}}_\mathrm{Field\
mode}\otimes\underbrace{\mathbb{I}\otimes\cdots\otimes
\mathbb{I}}_{i-1 \mathrm{\ terms}}\otimes\underbrace{\sigma^{x}}_{i
{th \
\mathrm{spin}}}\otimes\underbrace{\mathbb{I}\otimes\cdots\otimes\mathbb{I}}_{j-i-1
\mathrm{\ terms}}
\otimes\underbrace{\sigma^{y}}_\mathrm{j {th\ spin}}\otimes
\underbrace{\mathbb{I}\otimes\cdots\otimes \mathbb{I}}_{N-j \mathrm{\ terms}}.
\label{Eq:KronekerSpins}
 \end{eqnarray}
   
\section{The energy spectra for 4 spin chain}
\label{sec:examples4spins}
We diagonalize equation (\ref{Eq:HamOCOperator}), expressed in matrix form
using (\ref{Eq:KronekerField}-\ref{Eq:KronekerSpins}) for the case of
four spins, $N=4,$ to obtain the exact solutions for the energy
levels. To this end we solve the characteristic equation for $E$ i.e.
$$
\det\left(\widehat{H}-E\;\mathbb{I}_{32\times32}\right)=0.
$$
The Hamiltonian is a $32\times32$ matrix, and accordingly, there are 32
(some multiply degenerate) energy levels, which we found with help of
a computer algebra system.
%~\cite{Wolfram-Research:2014}.
For the open
spin chain, the results depend on whether the photon falls on the first
or the second spin, while the addition of the cyclic interactions term would appear
to make this distinction irrelevant for the closed chain, as the
results show as well. In order to check the computer derived formulae,
it is useful to compare the limit $J=0$ with the known solutions of
the two level Jaynes-Cummings model, which are $\pm G$ and $0$ in
units of $\hbar$. 

\subsection{Open chain -- edge coupled photon}
\begin{figure}[ht!]\centering
\includegraphics[width = 2.6in]{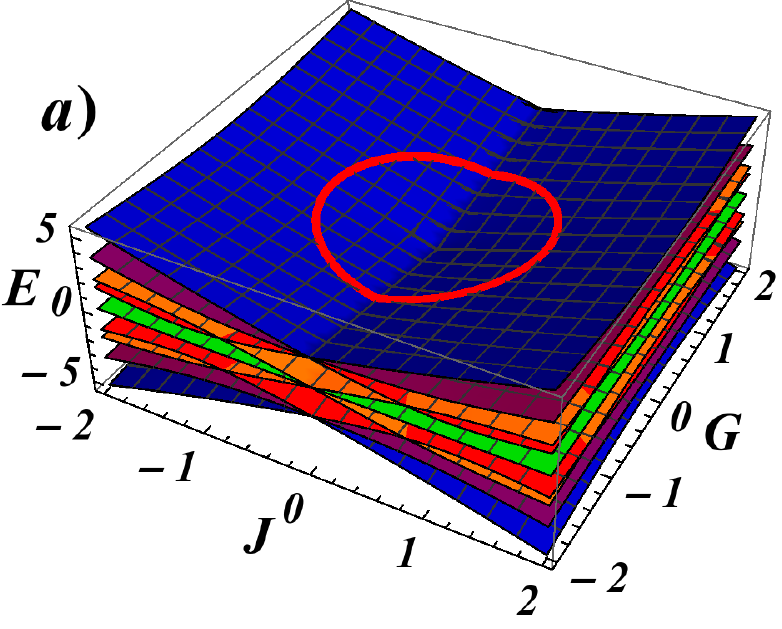}
\includegraphics[width = 3.3in]{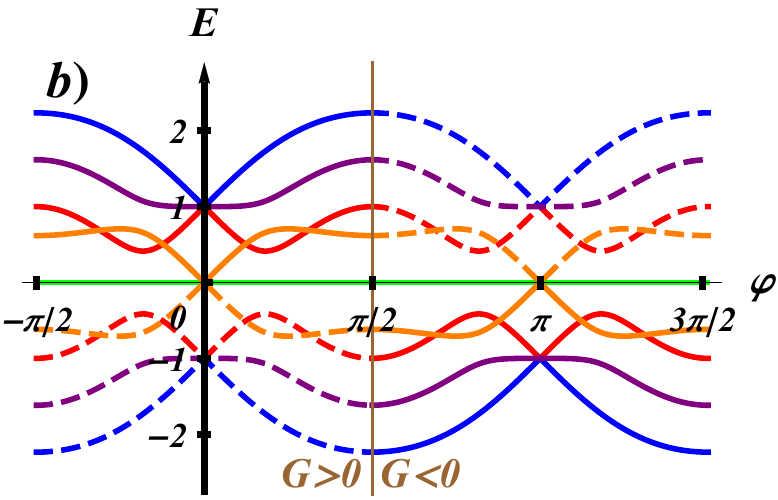}
\caption{(Color online) The energy spectra given by
equations~(\ref{Eq:EnSpectrumOpenFirst}) for a 4 spin chain coupled to
a photon at the edge of the chain. (a) Energy surfaces at various
finite strengths $G$ and $J.$ (b) The cross sections of the energy
surfaces at the red (solid line) circle shown in (a). See text for
further discussion of the notations and the symmetries of the
spectrum. Close to $\varphi = 0,$ $J\ll G$, and close to $
\varphi=\tfrac{\pi}{2},$
$G\ll J.$}
\label{Fig:EnSpectraOpenFirst}
\end{figure}

The spectrum in this case consists of 9 distinct energy branches shown
in Figure~\ref{Fig:EnSpectraOpenFirst}. The analytic expressions
, where the superscripts $D,Q,$ or
$O$ stand for the degeneracy of each level $2, 4,$ or $8$ times,
listed in decreasing order at $G=0$ are:
\begin{subequations}
\label{Eq:EnSpectrumOpenFirst}
\begin{align}
E^D_{1}(G,J)&=\sqrt{G^2+3 J^2+2 \sqrt{2 G^2 J^2+J^4}},\\
E^Q_{2}(G,J)&=\frac{\sqrt{G^2+3 J^2+\sqrt{G^4-2 G^2 J^2+5 J^4}}}{\sqrt{2}},\\
E^D_{3}(G,J)&=\sqrt{G^2+3 J^2-2 \sqrt{2 G^2 J^2+J^4}}, \\
E^Q_{4}(G,J)&=\frac{\sqrt{G^2+3 J^2-\sqrt{G^4-2 G^2 J^2+5 J^4}}}{\sqrt{2}},\\
E^O_{5}(G,J)&= 0,
\end{align}
\end{subequations}
and $E^Q_{6}= -E^Q_{4}, \;E^D_{7} = -E^D_{3}, \;E^Q_{8} = -E^Q_{2},
\;E^D_{9} = -E^D_{1}.$
The (at least) double degeneracy of the levels follows from the
Kramers theorem (we do not have an explicit magnetic field in
the Hamiltonian), where the photon acts as an 
additional spin in the chain for the purposes of the theorem
(see e.g. $\S$~60 of \cite{LandauLifshitzIII}).
To better show the behavior of the levels with predominantly
light-spin interaction ($J\ll G$) or predominantly spin-spin
interaction ($G\ll J$) we have used the substitution $G=C
\cos\varphi, J = C \sin\varphi,$ where in the
Figure~\ref{Fig:EnSpectraOpenFirst}(b), $C=1.$ With this notation,
when $\varphi \in (-\tfrac{\pi}{2}, 0)$ the chain is an AFM chain, and when
$\varphi \in (0, \tfrac{\pi}2)$ it is a FM chain. Additionally shown is the
region $\varphi > \tfrac{\pi}2,$ where $G$ formally takes negative values.
This would mean an overall change of the sign of the Hamiltonian, and
accordingly, we see an energy spectrum that consists of the same
levels, just arranged with opposite signs, so solid lines go to the
corresponding solid lines at the line at $\varphi = \tfrac{\pi}{2}$.

%%%%%fig.2%%%%%%%% 
%%%%%fig.2%%%%%%%%

\subsection{Open chain -- photon coupled to the second spin in the chain}
Due to the extra neighbor to the spin impacted by light, the
qualitative change in the spectrum, is that the quadruply degenerate
bands in (\ref{Eq:EnSpectrumOpenFirst}) split into doubly degenerate
bands, and thus all levels, except the zero energy, are now doubly
degenerate. The eigenvalues in explicit form look rather cumbersome,
so we first give the characteristic equation. It could be written in
factored form as:
\begin{align}
\det\left(\widehat{H}-E\;\mathbb{I}_{32\times32}\right)&= E^8 \left[E^4 - (G^2+3 J^2) E^2 + (G^2+J^2)J^2\right]^2 \nonumber\\
&\quad\times\left\{E^8 - (3G^2+9J^2)E^6+(3G^4+15G^2J^2+24J^4)E^4 \right. \nonumber \\
&\qquad- (G^6+7G^4J^2+23G^2J^4+21J^6)E^2 + G^6J^2+7G^4J^4\nonumber\\
&\qquad+\left. 3G^2J^6+5J^8\right\}^2 \nonumber \\
&=0.
\label{Eq:CharEqOpenSecond}
\end{align}

\begin{figure}[ht!]\centering
\includegraphics[width = 2.6in]{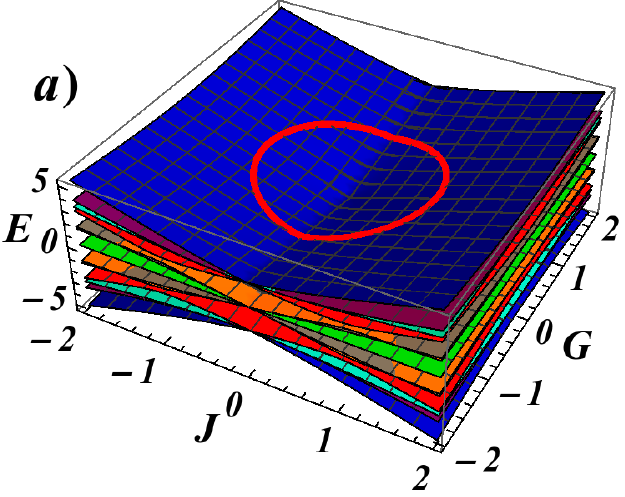}
\includegraphics[width = 3.3in]{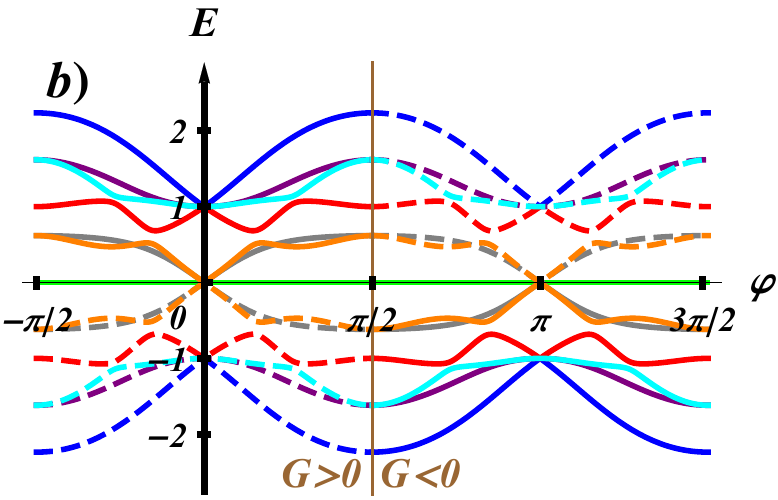}
\caption{(Color online) The energy spectra given by
equations~(\ref{Eq:CharEqOpenSecond}) for a 4 spin chain coupled to a
photon at second spin in the chain. Notations are similar to
Figure~\ref{Fig:EnSpectraOpenFirst}.}
\label{Fig:EnSpectraOpenSecond}
\end{figure}

The solutions of this equation are shown in
Figure~\ref{Fig:EnSpectraOpenSecond}, and there are 6
branches in the upper half-plane, with 13 overall distinct solutions.
From the first factor we get the octuply degenerate zero level, 
\begin{equation}E^O(G,J) =0,\end{equation}
from the second factor (expression in the square brakets) and the
third factor (curly brackets) the remaining doubly degenerated
eigenvalues. The expressions from the second factor:
\begin{equation}
E_{a,b}^D(G,J)= -E_{c,d}^D(G,J)
=\frac{\sqrt{G^2+3 J^2\pm\sqrt{G^4+2 G^2 J^2+5 J^4}}}{\sqrt{2}},
%\quad E_{c,d}^D(G,J) = -E_{a,b}^D(G,J), 
\end{equation}
look similar to the $E^Q_{2,4}(G,J)$ (and $E^Q_{6,8} (G,J)$) in
equation (\ref{Eq:EnSpectrumOpenFirst}), with the only change in sign
in the term in the second square root:  $-2 G^2 J^2 \rightarrow +2
G^2J^2.$ The expressions for those from the third factor (the eight
remaining lines on the figure), as solutions of fourth order equation
for $E^2,$ look involved and we only note that the topmost band, which
is given by one of the roots of this fourth order equation, looks
similar, with a slight decrease in the initial slope at $J\rightarrow
0,$ to the topmost band $E^D_{1}$ in~(\ref{Eq:EnSpectrumOpenFirst}).

%%%%%fig.3%%%%%%%%
%%%%%fig.3%%%%%%%%

\subsection{Closed chain}
\begin{figure}[ht!]\centering
\includegraphics[width = 2.4in]{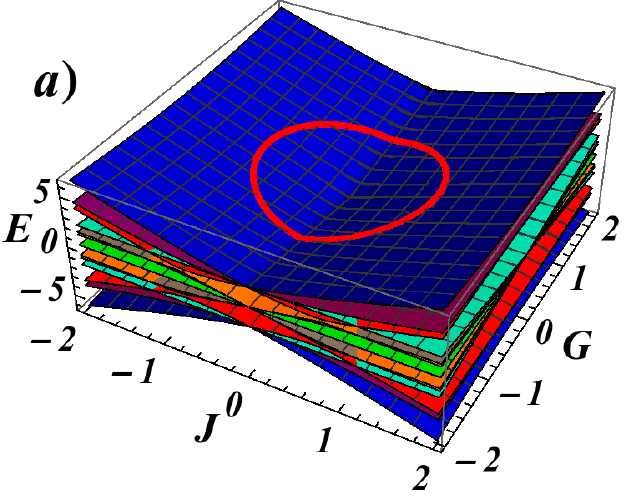}
\includegraphics[width = 3.1in]{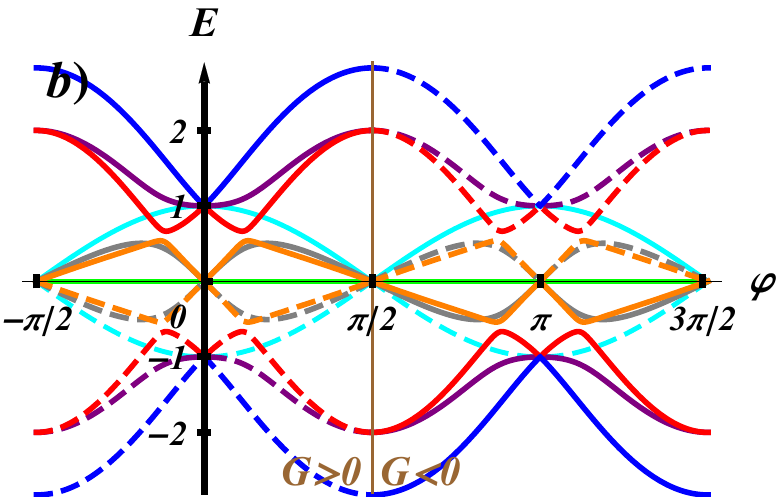}
\caption{(Color online) The energy spectra from
equation~(\ref{Eq:CharEqClosed}) for a 4 spin cyclic chain coupled to
a photon. Notations are similar to
Figure~\ref{Fig:EnSpectraOpenFirst}.}
\label{Fig:EnSpectraClosed}
\end{figure}

The 4-spin cyclic configuration is described by the following characteristic equation: 
\begin{align}
\det\left(\widehat{H}_{C}-E\;\mathbb{I}_{32\times32}\right)&=E^8 \left(E^2-G^2\right)^2 \left[E^4-(G^2+4J^2)E^2 +2 G^2 J^2\right]^2\nonumber\\
& \quad\times\left\{E^6-(2 G^2+12 J^2)E^4 +( G^4+6 G^2 J^2+32 J^4)E^2
-2 G^4J^2-8 G^2 J^4\right\}^2 \nonumber\\
&=0,
\label{Eq:CharEqClosed}      
\end{align}
whose 13 distinct solutions are shown in the Figure~\ref{Fig:EnSpectraClosed}. 
$\widehat{H}_{C}$ compared to~(\ref{Eq:HamOCOperator}), has an additional
spin-spin interaction term between the spins on sites $N$ and $1$;
thus closing the chain.
Except for the octuply degenerate zero level,
\begin{equation}E^O(G,J) =0,\end{equation}
the energy spectrum consists of doubly degenerate levels. The second factor gives: 
\begin{equation}
E_{A,B}^D(G,J)= \pm G,
\end{equation}
which is independent of $J$. In
Figure~\ref{Fig:EnSpectraClosed}(b) it appears as a cosine (cyan) line. We
have checked, that for a 3-spin chain there also is an energy level
that is linear in G, but has an additional, also linear, dependence on
$J.$ Here such levels come from a mixture of different basic spin
configurations, for which the magnetic part of the
Hamiltonian~(\ref{Eq:HamOCOperator}) produces canceling contributions.
The third, quadratic in $E^2,$ factor gives:
\begin{equation}
E^D_{a,b}(G,J)=-E^D_{c,d}(G,J)=\frac{\sqrt{G^2+4 J^2\pm \sqrt{G^4+16 J^4}}}{\sqrt{2}}.
%\textrm{ and } E^D_{c,d}(G,J)=-E^D_{a,b}(G,J).
\end{equation}

The last factor, curly brackets in~(\ref{Eq:CharEqClosed}), is cubic in $E^2$ and gives the
remaining 12 roots (counting the degeneracy), or 6 lines on the
figure, and the topmost band is one of these roots. The explicit
formulae are again somewhat cumbersome, and we skip writing them
explicitly.

%%%%%fig.4%%%%%%%%
%%%%%fig.4%%%%%%%%

\section{Conclusions}
\label{sec:conclusion}
In this report we have discussed some analytical results of applying
the Jaynes-Cummings model to an XY- AFM or FM spin $\tfrac12$ chain.
Because we limited ourselves to a single mode photon field, 
the resulting Hamiltonian in matrix form resembles the case of a pure spin chain 
with one extra spin site inducing anisotropy.
Nevertheless the light-spin interaction term does bring a novel
effect: it does not commute with the total spin 
$S_z^\mathrm{total} = \tfrac12\sum_{i=1}^{N} \sigma^z_i.$ This means that the eigenstates
of the XY-Jaynes-Cummings Hamiltonian consist of some mixtures of the product 
basis states, which have definite spin moment value. For that reason only for $G=0$ 
($\varphi=\tfrac{\pi}{2}$ point in figures~\ref{Fig:EnSpectraOpenFirst},
\ref{Fig:EnSpectraOpenSecond},
\ref{Fig:EnSpectraClosed}) it is possible to assign definite spin moment to the energy levels.
As can be seen from the equations, the 4-spin chain seems to be the
limit for the analytical formulae, since more sites would require
solving polynomial equations of more than 5-th power, which may not
have a closed form.
%With the numerical methods implemented in~\cite{Wolfram-Research:2014},
It was possible to obtain numerical
data for the energy levels of a chain made of  more than 4 spins, this would be
reported in some future work~\cite{futurework}.

\ack
This work was supported by EU FP7 INERA project grant agreement number
316309. We thank N.~Tonchev for bringing our attention to~\cite{RJZSHM2015}.

\section*{References}
\bibliographystyle{iopart-num}
\bibliography{INERA-lnn-2015-10-IOP}

\end{document}